\documentclass[a4paper,11pt]{article}
\usepackage{jcappub} 
\usepackage{lineno}
\usepackage{booktabs}
\usepackage{enumerate}
\usepackage{academicons}
\usepackage{xcolor}
\usepackage{fontawesome5}
\definecolor{orcidlogocol}{rgb}{0.65, 0.807, 0.223}

\title{\boldmath Impact of light sterile neutrinos on cosmological large scale structure}

\newcommand{\orcid}[1]{$\,$\href{https://orcid.org/#1}{\textcolor{orcidlogocol}{\faOrcid}}}

\author[a]{Rui Hu\orcid{0009-0008-1761-0717},}
\author[a]{Ming-chung Chu\orcid{0000-0002-1971-0403},}
\author[a]{Shek Yeung}
\author[a]{and Wangzheng Zhang\orcid{0000-0003-0102-1543}}

\affiliation[a]{Department of Physics, the Chinese University of Hong Kong, Sha Tin, Hong Kong SAR}

\emailAdd{1155168718@link.cuhk.edu.hk}

\abstract{Sterile neutrinos with masses on the $\mathrm{eV}$ scale are promising candidates to account for the origin of neutrino mass and the reactor neutrino anomalies. The mixing between sterile and active neutrinos in the early universe could result in a large abundance of relic sterile neutrinos, which depends on not only their physical mass $m_{\rm phy}$ but also their degree of thermalization, characterized by the extra effective number of relativistic degrees of freedom $\Delta N_{\rm eff}$. Using neutrino-involved N-body simulations, we investigate the effects of sterile neutrinos on the matter power spectrum, halo pairwise velocity, and halo mass and velocity functions. We find that the presence of sterile neutrinos suppress the matter power spectrum and halo mass and velocity functions, but enhance the halo pairwise velocity. We also provide fitting formulae to quantify these effects.}

\begin{document}
\maketitle
\flushbottom

\section{Introduction}
\label{sec:intro}
Over the past several decades, a comprehensive and varied program of experimental neutrino measurements has significantly advanced our understanding of the elusive neutrino sector. This extensive research has uncovered findings that suggest the existence of physics beyond the Standard Model (SM), particularly the presence of non-zero neutrino masses, as evidenced by the discovery of neutrino flavor mixing. Furthermore, experimental results indicate that all observed neutrinos (antineutrinos) exhibit left-handed (right-handed) chirality\footnote{In Goldhaber's experiment, the neutrinos' helicities are determined by the helicities of $\gamma$ ray emitted in the opposite direction of the neutrinos' \cite{Goldhaber:1958nb}, but since neutrinos are relativistic, the chirality is approximately equal to helicity.} within experimental uncertainty. The existence of right-handed neutrinos provides a natural explanation to the small masses of neutrinos \citep{Gariazzo:2015rra, Mohapatra:2005wk, Abazajian:2012ys}. They do not participate in weak interactions and only interact gravitationally, thus called sterile neutrinos. This paper focuses on light sterile neutrinos with masses in the eV range.

The existence of the eV-scale sterile neutrinos is also motivated by experimental neutrino anomalies. Two short-baseline neutrino oscillation experiments, \textit{LSND} \cite{LSND} and \textit{MiniBooNE} \cite{miniboone}, found an excess of electron neutrinos from a muon neutrino beam. A potential solution is to introduce an additional neutrino state with a mass of approximately $1\,\rm eV$. Several short-baseline reactor neutrino experiments, \textit{Daya Bay, MINOS+} and \textit{Bugey-3} \cite{Hu:2020uvx}, have established a stringent exclusion on the sterile neutrino parameter space for $|\Delta m_{41}^2|< 1.2\, \mathrm{eV}^2$. Additionally, the \textit{MiniBooNE} experiment \cite{abratenko2023first} also derives exclusion contours at the $95\%$ C.L. in the planes of mass-squared splitting $\Delta m_{41}^2$ and the sterile neutrino mixing angles $\theta_{e\mu}$ and $\theta_{ee}$, ruling out most of the parameter space suggested by the experimental anomalies. However, these experiments have not been able to impose a stringent exclusion for higher mass-squared splittings. 

In addition to terrestrial experiments, cosmology offers another effective method for measuring the properties of sterile neutrinos. Relic $\mathrm{eV}$-scale sterile neutrinos could be produced by neutrino oscillations in the early universe and behave like radiation. If the sterile neutrino state mixes sufficiently strongly with other neutrino states, there will be roughly equal numbers of sterile and SM neutrinos. Similar to active neutrinos, light sterile neutrinos delay the matter-radiation equality epoch, leaving distinct signatures on the cosmic microwave background (CMB). The number of relativistic degrees of freedom is usually parameterized by $N_{\rm eff}$, defined so that the relativistic energy density after electron-positron annihilation is given by 
\begin{equation}
    \rho_{\rm rad} = N_{\rm eff} \frac78 \left( \frac{4}{11} \right)^{4/3} \rho_\gamma,
\end{equation}
where $\rho_\gamma$ is the photon energy density. The standard cosmological model has $N_{\rm eff} \approx 3.046$, representing 3 types of active neutrinos\footnote{Currently, $N_{\rm eff}$ is suggested to be 3.044 \cite{Bennett:2020zkv, Froustey:2020mcq}; here we keep the old value for consistency, which will have minimal impact on the results.}, and $\Delta N_{\rm eff} \equiv N_{\rm eff} - 3.046$ denotes the extra relativistic degrees of freedom.  Assuming the sum of 3 active neutrino mass eigenvalues $\sum m_i = 0.06\, \rm eV$ with the normal hierarchy, the upper bounds $N_{\rm eff}< 3.29$ and $m_{\rm eff}<0.65\, \rm eV$ are obtained by fitting of Planck CMB and baryonic acoustic oscillation (BAO) data \cite{planck18}. Here, $m_{\rm eff} \equiv \Omega_{\nu_s} h^2 (94.1 \,\rm eV)$  is the effective mass of sterile neutrino, characterizing its cosmological energy density $\Omega_{\nu_s}$. The physical mass of a sterile neutrino is $m_{\rm phy} = (\Delta N_{\rm eff})^{-1} m_{\rm eff}$ when produced via the Dodelson mechanism \cite{Dodelson:1993je}. The cosmological constraint on $\rm eV$-scale sterile neutrinos can be relaxed by invoking non-standard cosmological production scenarios or introducing new neutrino interactions. For instance, by introducing the non-zero chemical potential of active neutrinos, BBN constraints allow $\Delta N_{\rm eff}<0.52$ \cite{neutrinoxi}. Moreover, taking $\Delta N_{\rm eff}$ to be $0.29$ from CMB constraints, the mass bound becomes $m_{\rm phy}<2.24\,\rm eV$. Hence, the $\rm eV$-scale light sterile neutrinos cannot be completely ruled out. The degree of sterile neutrino thermalization quantified by $\Delta N_{\rm eff}$ plays an important role in BBN and CMB constraints of sterile neutrino parameters.  

In a later epoch, relic sterile neutrinos become non-relativistic. Nevertheless, sterile neutrinos stream freely, preventing them from clustering at scales smaller than their free-streaming length \cite{Lesgourgues:2013sjj}. Structures below this range tend to be washed out due to the high velocity dispersion of sterile neutrinos, which imprints a distinct signature on large-scale structure (LSS). Combining local LSS surveys,  such as galaxy shear power spectra from CFTHLenS surveys \cite{Miller_2013}, SDSS Ly-$\alpha$ forest data \cite{SSDS}, and X-ray \textit{Chandra} observations \cite{Vikhlinin_2009}, the constraints on sterile neutrino parameters are $m_{\rm eff}<0.22\,\rm eV$ and $\Delta N_{\rm eff}< 1.11$ (95\% C.L.) \cite{Costanzi_2014}.

Although combining Planck CMB data with future LSS data will not improve the constraint on $\Delta N_{\rm eff}$, it will significantly increase the sensitivity to $m_{\rm eff}$. Combining CMB data with galaxy and weak lensing data from Euclid \cite{Amendola2018} and DESI \cite{levi2013desi}, the sensitivity could reach $m_{\rm eff}< 60\,\rm meV$. Furthermore, future CMB experiments, such as CORE-M5, are expected to reach $m_{\rm eff}< 37\,\rm meV$ and $\Delta N_{\rm eff} < 0.053$ \cite{Valentino_2018}.

An effective way to incorporate sterile neutrinos into N-body simulations is to treat them as an additional type of collisionless particles \cite{Nascimento:2021wwz}. In addition to the bulk velocity, the sterile neutrino particles receive a thermal velocity from the Fermi-Dirac distribution. The same method can be used generally for hot dark matter (HDM). However, the resolution of sterile neutrinos or HDM particles should be at least twice of that CDM particles, which is computationally demanding. Another method to simulate sterile neutrinos uses a grid-based approach, treating sterile neutrinos as a linearly evolving background density \citep{Ali12, Chen:2020kxi}. In this linear response approach (LRA), the sterile neutrino evolution is calculated semi-analytically, accounting for their interaction with the nonlinear dark matter field. This LRA makes the neutrino-involved simulation faster than the particle approach with the same accuracy. In this article, we construct our sterile neutrino-involved N-body simulation following the LRA approach. 

Armed with high-resolution sterile neutrino-involved N-body simulations, one can investigate the impact of sterile neutrinos on various cosmological observables. For instance, it is well-known that light neutrinos suppress the matter power spectrum by $1-8\Omega_\nu/\Omega_m$, in a range of wave numbers $k$, giving rise to a spoon-shaped matter power spectrum due to free-streaming effects \cite{Agarwal:2010mt}. Here, $\Omega_\nu$ and $\Omega_m$ denote the dimensionless energy densities of neutrinos and total matter, respectively. Sterile neutrinos may induce similar effects. The two-point correlation function (2PCF) is also an old but powerful tool for studying galaxy and matter clustering across different scales. It has contributed to significant discoveries, such as baryonic acoustic oscillations (BAO) \cite{BAO}, which serve as a reliable standard ruler due to their robustness against systematics \cite{Ross_2016}.  Furthermore, 2PCF can be used to constrain various cosmological parameters \citep{jing1998spatial, Cole_2005, Zhai_2023}.

The matter mean pairwise peculiar velocity, or pairwise velocity in short, denoted as $v_{12}(r)$, has also been used to constrain cosmological parameters, such as $\Omega_m$ and $\sigma_8$ \citep{Ma_2015, Juszkiewicz2000, Feldman_2003, Zhang:2024pyf},  modified gravity \cite{MG}, and kinematic Sunyaev-Zeldovich (kSZ) effects \cite{Bhattacharya_2007}. Both the pairwise velocity and its dispersion have been well studied \citep{sheth1996distribution, sheth2001linear, jing1998spatial,mo1993pairwise}. Besides, it can also be used to measure the neutrino mass and asymmetry \cite{Zhang:2023otn}.

The free-streaming effect of warm dark matter can also be revealed in halo statistics, such as the halo mass and velocity functions. The halo mass function (HMF) is sensitive to various cosmological parameters and baryonic physics \cite{cui2012effects, stanek2009effect}. It is also useful for studying the impacts of massive neutrinos \cite{castorina2014cosmology}. Similarly, the maximum circular velocity (MCV) of halos could also be related to the HMF \cite{kochanek2001dynamical} and the halo-galaxy connection \cite{zehavi2019prospect, gonzalez2000velocity}. 

The article is organized as follows. Section \ref{sec: aspects} briefly reviews relic sterile neutrinos and elucidates our sterile neutrino-involved simulations. Section \ref{sec:sim} presents convergence tests of our simulations and the cosmological parameters used. Section \ref{sec:results} discusses the impact of sterile neutrinos on various cosmology observables. We summarize our findings in Section \ref{sec:con}. Appendices A, B, and C present the overview of our sterile neutrino-involved simulation method, the Markov-Chain Monte Carlo (MCMC) refitting results of cosmological parameters, and detailed tests, respectively. 

\section{Theoretical and numerical aspects}
\label{sec: aspects}
\subsection{Cosmic sterile neutrino background}
We assume that the cosmological sterile neutrinos maintain the same temperature as that of the relic standard model (SM) neutrinos, $T_\nu$. The energy density of these SM neutrinos is 
\begin{equation}
\label{eq:active_intergration_1}
    \rho_{\nu_a}(T_\nu) = 2\times \frac{1}{2\pi^2} \sum_{i=1}^3 \int_0^\infty \frac{E_i}{\exp(E_i/T_\nu) + 1} p^2 \mathrm{d}p, 
\end{equation}
where $E_i = \sqrt{p^2 + m_i^2}$ is the energy for SM neutrinos, with $m_i$ the mass eigenvalues. The factor of 2 accounts for the antiparticles. Cosmic neutrinos decouple from the thermal bath at temperature $T_\nu\sim 1\, \mathrm{MeV}$, and the distribution has been frozen since then. Since neutrinos are still highly relativistic at $T_\nu\sim 1\, \mathrm{MeV}$, the $E_i$ in the denominator in Eq.~(\ref{eq:active_intergration_1}) can be simply replaced by the momentum $p$: 
\begin{equation}
    \label{eq:active_intergration_2}
    \rho_{\nu_a}(T_\nu) = 2\times \frac{1}{2\pi^2} \sum_{i=1}^3 \int_0^\infty \frac{E_i}{\exp(p/T_\nu) + 1} p^2 \mathrm{d}p.
\end{equation}
Ref.~\cite{Gariazzo_2019} considered a $3+1$ active-sterile neutrino system in the early universe with a full mixing matrix. The sterile neutrinos are not fully thermalized, but their distribution function follows 
\begin{equation}
\label{eq:nonfully}
    f_{\nu_s}^0 = \frac{1}{2\pi^2} \frac{\Delta N_{\rm eff}}{\exp(p/T_\nu) + 1}.
\end{equation}

\subsection{Cosmological background evolution}
We start with the deviation of the Hubble expansion history from that of the standard $\Lambda$CDM universe. The Hubble expansion rate is determined by the Friedmann equation including neutrinos,
\begin{equation}
\label{eq:origin H evolution}
    H^2(a) \equiv \left(\frac{\dot{a}}{a}\right)^2 = H_0^2 (\Omega_{\gamma,0} a^{-4} + \Omega_{cb,0} a^{-3} + \Omega_\Lambda + \Omega_{\nu_a}(a) + \Omega_{\nu_s}(a)),
\end{equation}
where $\Omega_{\gamma,0}$, $\Omega_{cb,0}$, and $\Omega_\Lambda$ are the current energy densities of photons, cold dark matter plus baryonic matter, and dark energy, respectively, and $H_0=H(a=1)$. The evolution of active and sterile neutrino energy densities, $\Omega_{\nu_a}$ and $\Omega_{\nu_s}$, can be calculated using their distribution functions, such as Eq.~(\ref{eq:nonfully}).

\subsection{Sterile neutrino free-streaming effect}
After decoupling, neutrinos stream freely. Similar to active neutrinos, cosmological sterile neutrinos are only weakly bound by the gravitational potential below the free-streaming scale because of their high thermal speeds. When sterile neutrinos become non-relativistic during matter domination, the free-streaming length is given by \cite{Lesgourgues:2013sjj}
\begin{equation}
\begin{aligned}
    &\lambda_{\rm fs} \approx 8.10 (1+z) \frac{H_0}{H(z)} \left(\frac{1\,\rm{eV}}{m_{\rm phy}}\right) h^{-1}\, \mathrm{Mpc}.
\end{aligned}
\end{equation}
Due to free-streaming effects, the growth of $3+1$ neutrino over-density $\delta_\nu$ is much slower than that of CDM plus baryonic matter $\delta_{cb}$, which leads to a smaller total matter over-density field $\delta_{\rm t}$ than that in the $\Lambda$CDM model due to the averaging effect: 
\begin{equation}
    \delta_{\rm t} = (1-f_\nu) \delta_{cb} + f_\nu \delta_\nu = (1-f_\nu) \delta_{cb} + f_{\nu_a} \delta_{\nu_a} + f_{\nu_s}\delta_{\nu_s},
\end{equation}
where $f_\nu = (\Omega_{\nu_a}+\Omega_{\nu_s}) / \Omega_m = f_{\nu_a} + f_{\nu_s}$ is the ratio between energy densities of $3+1$ neutrinos and total matter. $\delta_{\nu_a}$ ($\delta_{\nu_s}$) is the active (sterile) neutrino over-density.  Hence, in cosmological simulations, the total matter density field can be written as 
\begin{equation}
\label{eq:correction of density}
    \tilde{\rho}_{\rm t}(\mathbf{k}) = \tilde{\rho}_{\rm cb} (\mathbf{k}) + \tilde{\rho}_{\rm cb} (\mathbf{k}) \frac{f_\nu}{1 - f_\nu} \times \frac{\delta_\nu}{\delta_{\rm cb}}, \quad \frac{\delta_\nu}{\delta_{\rm cb}} \approx \left(\frac{P_\nu(\mathrm{k})}{P_{\rm cb}(\mathrm{k})}\right)^{1/2},
\end{equation}
where $P_\nu(\mathrm{k})$ and $P_{\rm cb}(\mathrm{k})$ are the $3+1$ neutrino power spectrum and CDM-baryon power spectrum, respectively. Similar to that for active neutrinos \cite{Zeng:2018pcv}, the linear growth equation of the sterile neutrino over-density field is 
\begin{equation}
\label{eq:linear evolution}
\begin{aligned}
    \tilde{\delta}_{\nu_s} = \tilde{\delta}_{\nu_s}(s_i, \mathbf{k})\Phi\left[\mathbf{k}(s-s_i)\right] \left[1 + (s - s_i)a_i^2 H(a_i)\right] + 4\pi G \int_{s_i}^s a^4 (s-s^\prime) \Phi\left[ \mathbf{k}(s-s^\prime)\right] \\ 
    \times \left[ \bar{\rho}_{cb}(s^\prime) \delta_{cb} (s^\prime, \mathbf{k}) + \sum_{i} \bar{\rho}_{\nu_i}(s^\prime) \delta_{\nu_i} (s^\prime, \mathbf{k}) + \bar{\rho}_{\nu_s}(s^\prime) \delta_{\nu_s} (s^\prime, \mathbf{k}) \right] \mathrm{d}s^\prime,
\end{aligned}
\end{equation}
where $s$ is the comoving time defined in Eq.~(\ref{eq: coordinates}), $\tilde{\delta}_{\nu_i}$ and $\tilde{\delta}_{\nu_s}$ are over-density fields of the $3$ active neutrinos and sterile neutrinos in $k$-space, respectively, and
\begin{equation}
\label{eq:phi}
    \Phi(\mathbf{q}) \equiv \frac{\int f_{\nu_s}^0(\mathrm{p}) e^{-i\mathbf{q\cdot p}}\mathrm{d^3 p}}{\int f_{\nu_s}^0(\mathrm{p})\mathrm{d^3 p}}.
\end{equation}
The derivation of Eqs.~(\ref{eq:linear evolution}) and (\ref{eq:phi}) is shown in Appendix~\ref{Appen: evolution}. The grid-based sterile neutrino-involved N-body simulations are similar to the method presented in \cite{Zeng:2018pcv}.  Appendix~\ref{append:simulation} overviews the details of the simulation procedure. 

\section{Neutrino-involved N-body simulations}
\label{sec:sim}
\subsection{Standard test}
\label{sec:test}
\begin{figure}[tpb]
    \centering
    \includegraphics[width=0.9\linewidth]{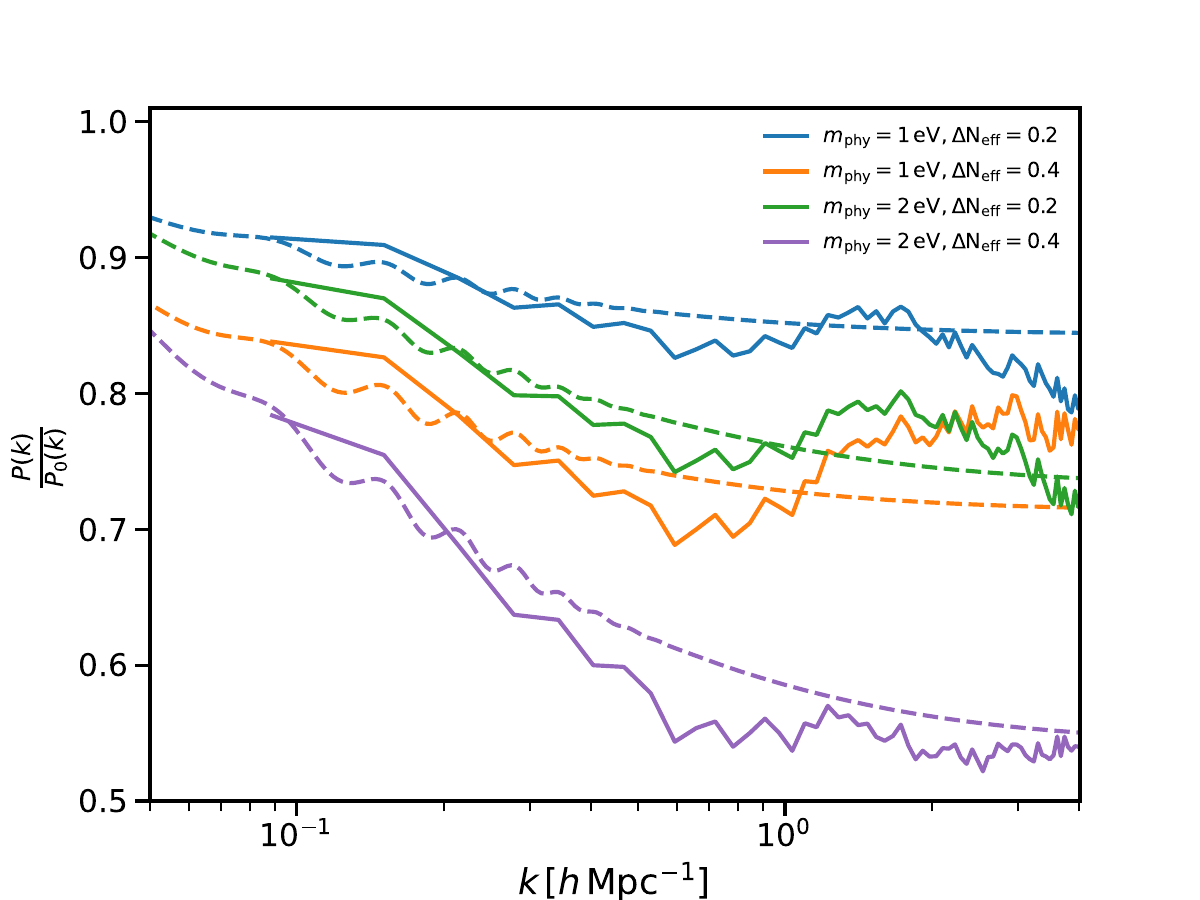}
    \caption{Ratios of total matter power spectra for T1, T2, T3, and T4 to that of T0 at redshift $z=0$, in linear theory (dashed lines) and simulations (solid lines).}
    \label{fig:suppression}
\end{figure}

 In our simulations, active neutrinos with 3 degenerate masses are considered with total mass $\sum m_{\nu_i}=0.06\, \rm eV$. We have modified \texttt{Gadget-2} \cite{Springel_2005} to incorporate both active and sterile neutrinos. The initial conditions are generated by \texttt{2LPTic} \cite{Crocce_2006}, incorporating the active and sterile neutrino effects in the Friedmann equation. The initial power spectrum of CDM plus baryons and $3+1$ neutrinos at $z=99$ are calculated using \texttt{CAMB}\footnote{https://camb.info/} \cite{Lewis_2000}. 
\begin{table}[tpb]
    \centering
    \begin{tabular}{ccllll} \toprule
         &  T0 & T1& T2 & T3 & T4\\ \midrule 
         $m_{\rm phy} [\rm eV]$& 
    0 &1& 1& 2&2\\ 
 $\Delta N_{\rm eff}$& 0& 0.2& 0.4& 0.2&0.4\\
 $m_{\rm eff} [\rm eV]$& 0& 0.2& 0.4& 0.4&0.8\\ \bottomrule
    \end{tabular}
    \caption{Combinations of $m_{\rm phy}$ and $\Delta N_{\rm eff}$ with fixed cosmological parameters. Here, $m_{\rm phy}$ is the physical mass of the sterile neutrinos and $m_{\rm eff} \equiv m_{\rm phy}\cdot \Delta N_{\rm eff}$ is the corresponding effective mass.}
    \label{tab:parameter with fixed}
\end{table}
For validation of the simulation, we studied 5 cases, with different combinations of $m_{\rm phy}$ and $\Delta N_{\rm eff}$ shown in Table~\ref{tab:parameter with fixed}. The background cosmology is fixed using Planck18 best-fit cosmological parameters \cite{planck18} so that the total matter density and baryon density are the same. The active and sterile neutrino densities are subtracted from the CDM density.  The simulations are conducted within a boxsize of $100 h^{-1}\,\mathrm{Mpc}$, comprising $N_{\rm CDM} = 128^3$ CDM particles. For the convergence test, T1 is run with $N_{\rm CDM} = 128^3, 256^3, 512^3$. 
\begin{figure}[tpb]
    \centering
    \includegraphics[width=0.9\linewidth]{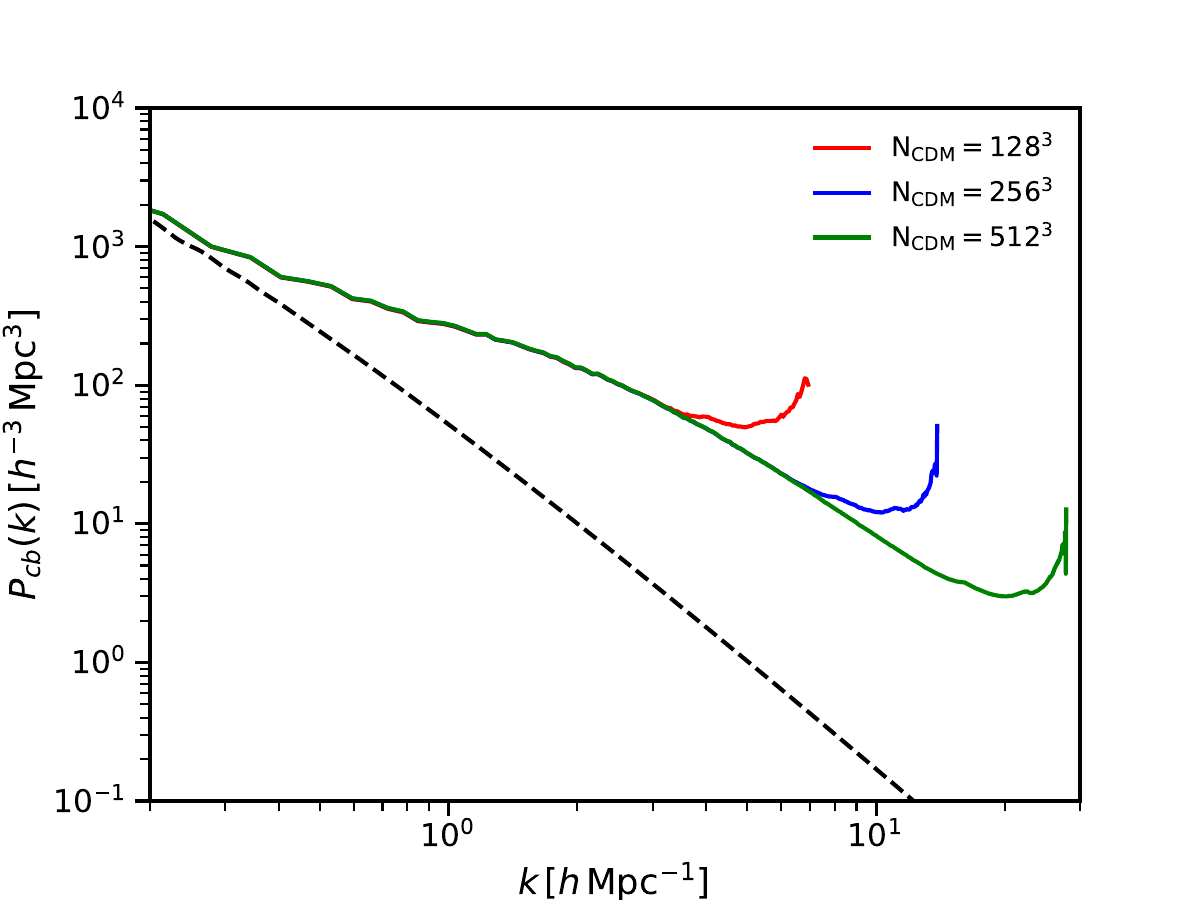}
    \caption{Power spectra of CDM plus baryons with different resolutions at redshift $z=0$. The solid lines with different colors show the simulation results with different resolutions. The dashed black line presents the linear power spectrum of CDM plus baryons for reference.}
    \label{fig: test_resolution}
\end{figure}
As shown in Figure~\ref{fig: test_resolution}, the results of neutrino-involved simulations converge well. With a fixed cosmology, the power spectrum of CDM plus baryons has a spoon-shape suppression at $k\sim 1 h\,\mathrm{Mpc}^{-1}$ impacted by sterile neutrinos as seen in Figure~\ref{fig:suppression}, which is similar to the effect seen with active neutrinos. 

\subsection{Consistent simulations}
To study the effects of sterile neutrinos on cosmic structure evolution, we use the same combinations of $m_{\rm phy}$ and $\Delta N_{\rm eff}$ in Table~\ref{tab:parameter with fixed}, but the other cosmological parameters are derived via refitting, utilizing the \texttt{Cosmomc} \footnote{https://cosmologist.info/cosmomc/} software \cite{Lewis_2002}, of the Planck 2018 plikHM\_TTTEEE and BAO data \cite{planck18, BAO}. Detailed cosmological parameters are shown in Table~\ref{tab:parameter by refitting}. $N_{\rm CDM}=1024^3$ and boxsize $L=1000\,h^{-1}\, \mathrm{Mpc}$ are used. Dark matter halos are identified using the \texttt{Rockstar} \footnote{https://bitbucket.org/gfcstanford/rockstar/src/main/}  software \cite{Behroozi_2013}, which employs adaptive hierarchical refinement of friends-of-friends groups in phase-space dimensions, and analyses are conducted across multiple redshifts.
\begin{table}[tpb]
    \centering
    \begin{tabular}{cccccclll} 
        \toprule
         &  $m_{\rm phy} [\rm eV]$& $\Delta N_{\rm eff}$& $H_0[\mathrm{km\,s^{-1} Mpc^{-1}}]$& $\Omega_c h^2$& $\Omega_b h^2$&$\Omega_\Lambda$ &$n_s$ &$\ln{A_s}$\\ \midrule
 A0&  0& 0& 67.75& 0.1193& 0.0224& 0.6897& 0.967&3.046\\ 
 B1& 1& 0.2& 67.90& 0.1215& 0.0226& 0.6814& 0.973&3.062\\
 B2& 1& 0.4& 68.11& 0.1236& 0.0228& 0.6739& 0.979&3.078\\
 C1& 2& 0.2& 67.46& 0.1199& 0.0226& 0.6760& 0.970&3.063\\
 C2& 2& 0.4& 67.27& 0.1204& 0.0228& 0.6631& 0.974&3.082\\ \bottomrule
    \end{tabular}
    \caption{Combinations of $m_{\rm phy}$ and $\Delta N_{\rm eff}$ and the corresponding refitted cosmological parameters.}
    \label{tab:parameter by refitting}
\end{table}

\section{Results}
\label{sec:results}
\subsection{Sterile neutrino effects on matter power spectrum and two-point correlation function}
\label{subsec:ps}
\begin{figure}[tpb]
    \centering
    \includegraphics[width=0.9\linewidth]{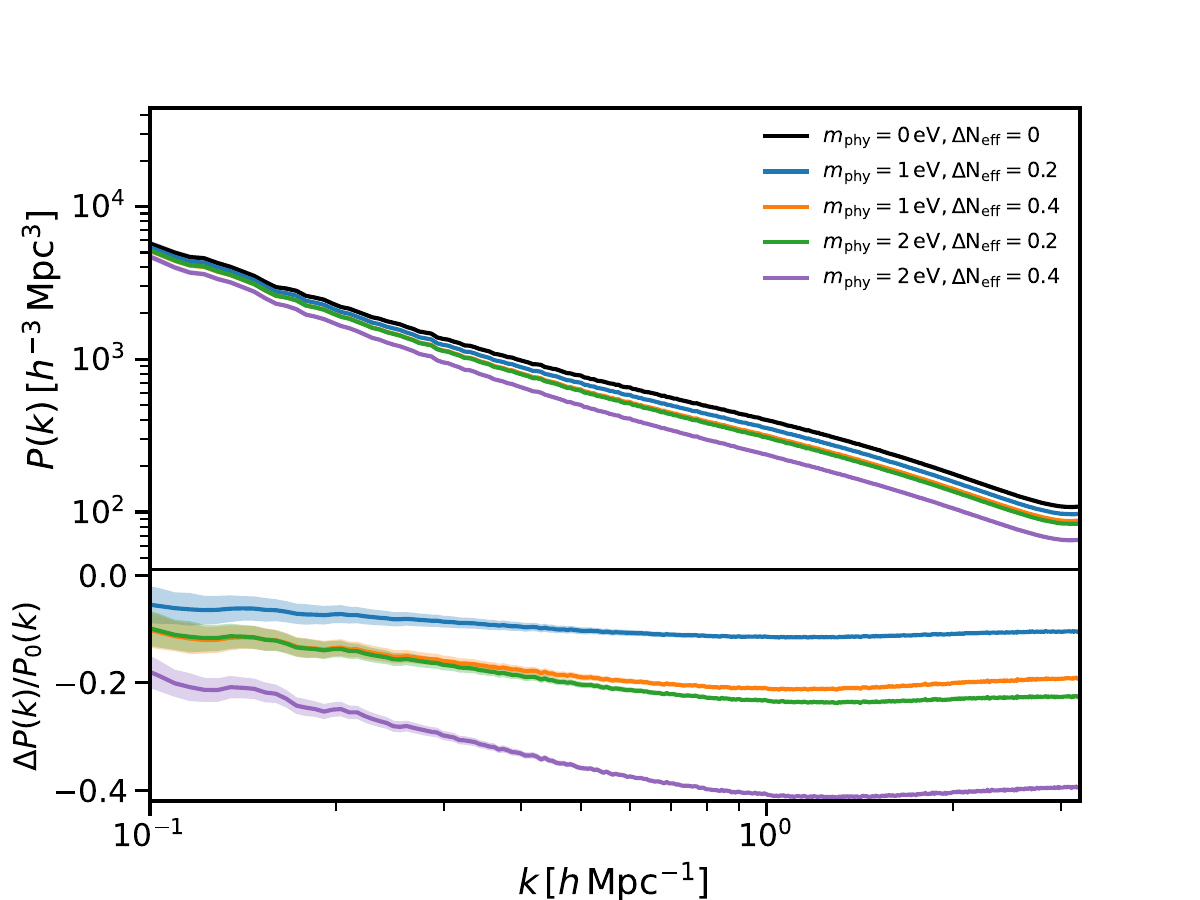}
    \caption{Total matter power spectra (upper panel) and fractional deviations (lower panel) with respect to the fiducial case A0 for different combinations of $m_{\rm phy}$ and $\Delta N_{\rm eff}$. Shadowed regions show the statistical errors.}
    \label{fig:pk_effect}
\end{figure}
\begin{figure}[tpb]
    \centering
    \includegraphics[width=0.9\linewidth]{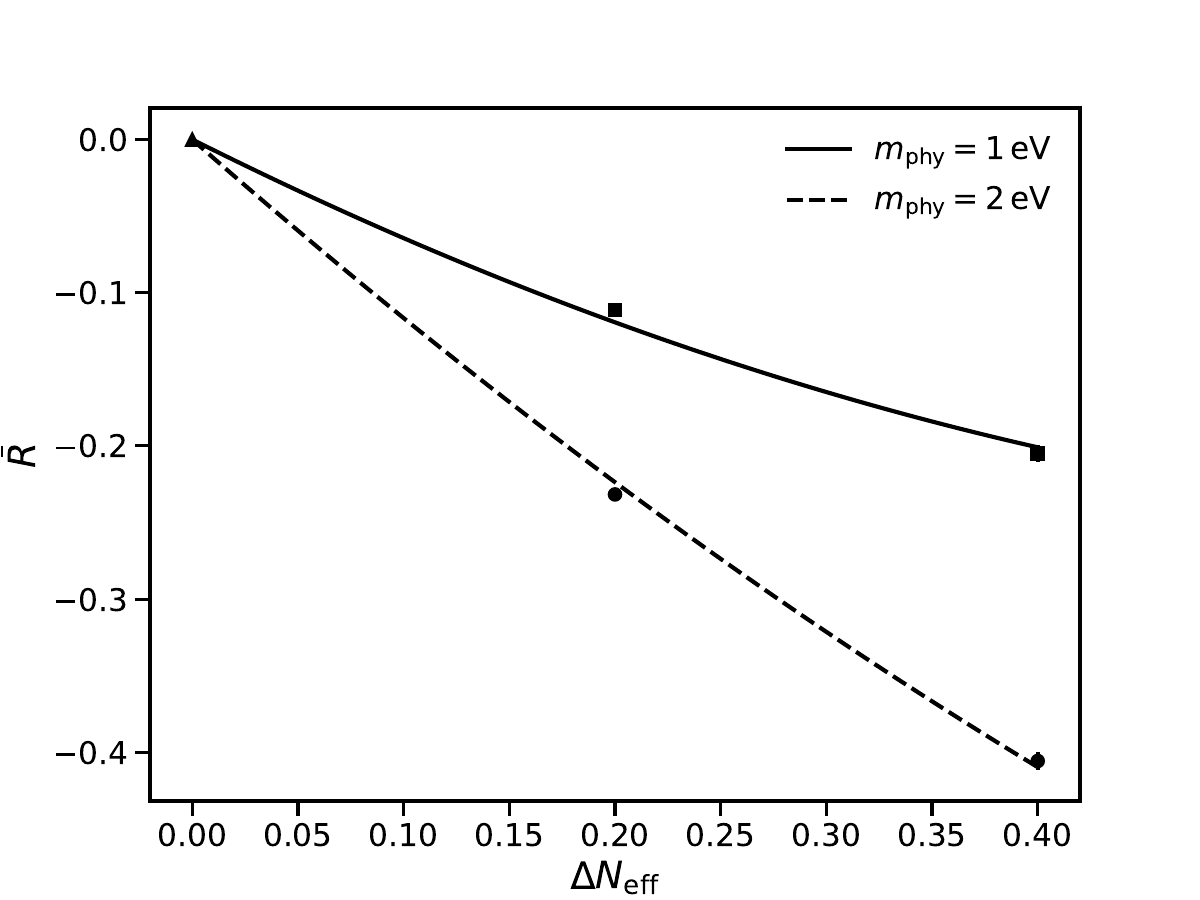}
    \caption{Fractional deviations of the matter power spectrum from that of A0 (triangle) averaged over the range $k\in[0.7,2.5]h\,\rm{Mpc}^{-1}$ for $m_{\rm phy} = 1\,\rm{eV}$ (squares) and $2\,\rm{eV}$ (circles). The dashed and solid lines are the fitting results.}
    \label{fig:lin_regression}
\end{figure}

We analyze the matter power spectrum utilizing the \texttt{Pylians} software \cite{Pylians}. Figure~\ref{fig:pk_effect} shows the total matter power spectra and their fractional deviations from that of the fiducial case (A0) for different combinations of $m_{\rm phy}$ and $\Delta N_{\rm eff}$,
\begin{equation}
    R(k) = \Delta P(k)/P_0(k) = \frac{P(k) - P_0(k)}{P_0(k)}, 
\end{equation}
showing a suppression of around 10\% already for $m_{\rm phy}=1\,\rm{eV}$ with $\Delta N_{\rm eff} =0.2$. Although increasing $m_{\rm phy}$ or $\Delta N_{\rm eff}$ will suppress the matter power spectrum, their degeneracy cannot be represented by $m_{\rm eff}$ alone. For example, even though the combinations B2 and C1 have the same  $m_{\rm eff}=m_{\rm phy}\cdot\Delta N_{\rm eff}$, their total matter power spectra differ by a few percent. To quantify the effects of sterile neutrinos, we take $R(k)$ averaged over the range of $[0.7,2.5]  h\,\rm{Mpc}^{-1}$,  yielding
\begin{equation}
    \bar{R} = \frac{\sum R(k)}{N_{\rm kbin}},
\end{equation}
where $N_{\rm kbin}$ is the number of $k_i$ bins sampled in this range. $\bar{R}$ can be fitted by expanding $\bar{R}(m_{\rm phy}, \Delta N_{\rm eff})$ at the fiducial values $m_{\rm phy}=0, \Delta N_{\rm eff}=0$ (A0) and keeping the lowest order terms: 
\begin{equation}
    \begin{aligned}
        \bar{R}(m_{\rm phy}, \Delta N_{\rm eff}) = C_{mm} \cdot \left(\frac{m_{\rm phy}}{1\,\rm{eV}}\right)^2 + C_{nn} \cdot (\Delta & N_{\rm eff})^2 + C_{mn} \cdot \frac{m_{\rm phy}}{1\,\rm{eV}} \times \Delta N_{\rm eff} \\ 
        & + C_{m} \cdot m_{\rm phy} + C_n \cdot \Delta N_{\rm eff}, 
    \end{aligned}
\end{equation}
where $C_{mm}, C_{nn}, C_{mn}, C_m$ and $C_n$ are expansion coefficients. When $\Delta N_{\rm eff}$ is equal to $0$, there is no sterile neutrinos effects on the matter power spectrum, so $\bar{R}(m_{\rm phy}, \Delta N_{\rm eff}=0)=0$. Therefore, the fitting formula can be simplified as:
\begin{equation}
    \bar{R}(m_{\rm phy}, \Delta N_{\rm eff}) = C_{nn} \cdot (\Delta N_{\rm eff})^2 + C_{mn} \cdot \frac{m_{\rm phy}}{1\,\rm{eV}} \times \Delta N_{\rm eff} + C_n \cdot \Delta N_{\rm eff}.
\end{equation}
The fitted parameters are $C_{nn} = 0.472 \pm 0.177$, $C_{mn} = -0.522 \pm 0.028$, and $C_n = -0.169 \pm 0.078$. The results are shown in Figure~\ref{fig:lin_regression}.

Since many catalogs list the positions of galaxies or galaxy clusters, it is useful to study the matter spatial distribution. Besides the matter power spectrum, we also calculate the two-point correlation function,
\begin{equation}
    \xi(r) = \langle \delta_{cb}(\mathbf{r}^\prime) \delta_{cb}(\mathbf{r^\prime + r}) \rangle.
\end{equation}
One can estimate $\xi(r)$ through the pair counts of particle points and random points \cite{Landy_xi},
\begin{equation}
    \xi(r) = \frac{DD-2DR + RR}{RR},
\end{equation}
where $DD$, $DR$, and $RR$ are the normalized numbers of particle-particle pairs, particle-random pairs, and random-random pairs, respectively.  A simple estimate of the statistical error assumes that the number of independent pairs in a given $r$ bin follows a Poisson distribution.

\begin{figure}[tpb]
    \centering
    \includegraphics[width=0.9\linewidth]{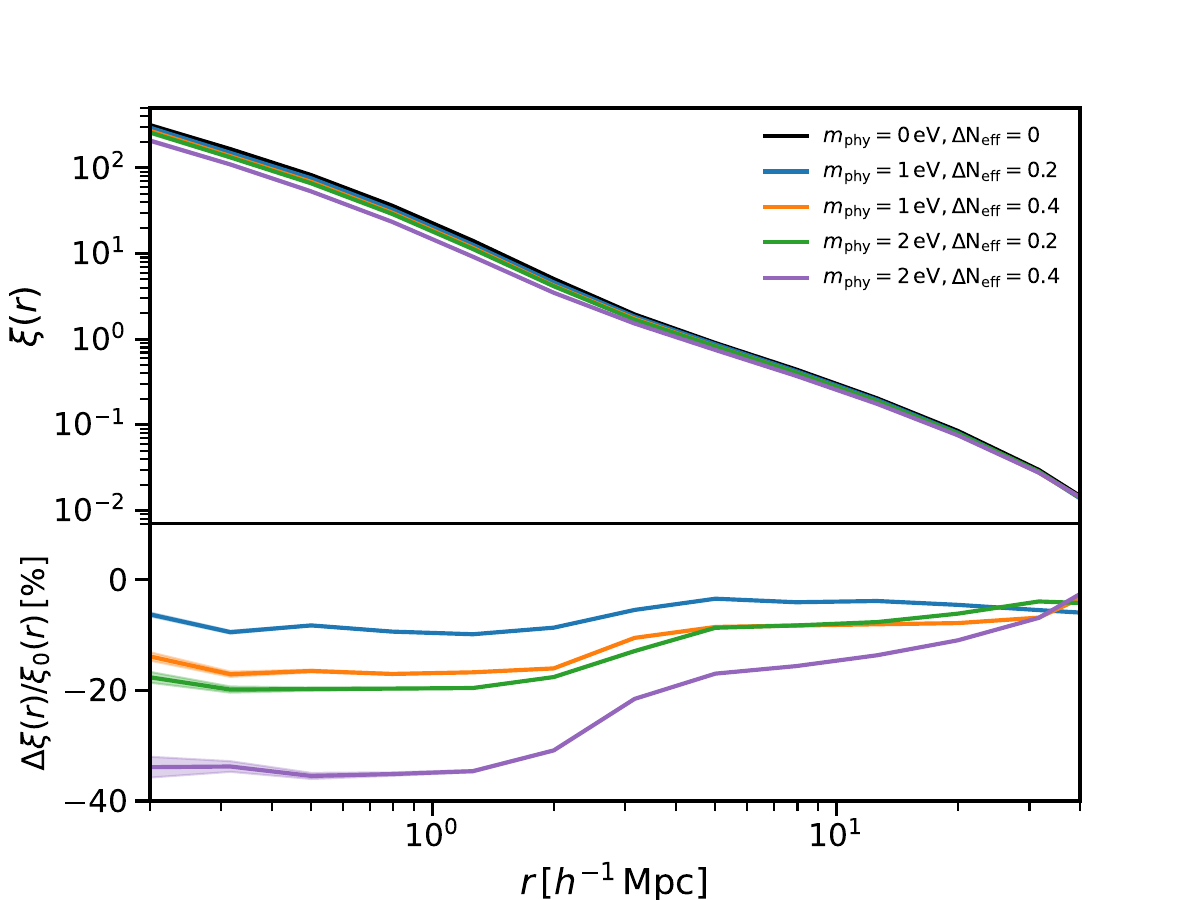}
    \caption{Same as Figure~\ref{fig:pk_effect}, but for the two-point correlation function.}
    \label{fig:xi_ratio}
\end{figure}

Figure~\ref{fig:xi_ratio} shows that $\xi(r)$ is suppressed for increasing $m_{\rm phy}$ or $\Delta N_{\rm eff}$  by up to around 30\% in the case of C2. The degeneracy of the two parameters is broken at $r < 4\,h^{-1}\, \mathrm{Mpc}$, which is consistent with the results from the matter power spectrum.

\subsection{Sterile neutrino effects on pairwise velocity}
\label{subsec:ppwv}
The velocity field also carries half of the phase-space information of cosmological structures. In this section, we focus on the pairwise velocity and its dispersion. The pairwise velocity $v_{12}$ is defined as the peculiar velocity difference of a pair of objects projected along their line of separation, averaged over all pairs at separation $r$ \cite{Ferreira1999},
\begin{equation}
    v_{12}(r) \equiv \langle \left(\mathbf{v}_1 - \mathbf{v}_2\right)\cdot \mathbf{\hat{r}} \rangle,
\end{equation}
where $\mathbf{\hat{r}} =(\mathbf{r}_1 - \mathbf{r}_2)/\left| \mathbf{r}_1 - \mathbf{r}_2 \right|$ and $\mathbf{v}_{1,2}$ is the peculiar velocity of object 1 and 2, respectively.  The pairwise velocity $v_{12}$ shows the tendency of two objects to move closer ($v_{12}< 0$) or farther apart ($v_{12}>0$) from each other. The pairwise velocity dispersion
\begin{equation}
    \sigma_{12}(r) = \langle \left[ (\mathbf{v}_1 - \mathbf{v}_2 ) \cdot \mathbf{\hat{r}} \right]^2 \rangle^{1/2}
\end{equation}
describes the r.m.s. fluctuations of $v_{12}$. Both $v_{12}$ and $\sigma_{12}$ help us understand how objects such as galaxies move relative to each other due to their gravitational interactions.

The pairwise velocity can be estimated analytically. For an evolving point distribution, the averaged number of neighbors within a comoving distance $r$ of an object is 
\begin{equation}
\label{eq:neighbors}
    N(r, t) = 4\pi \bar{n}(t) a^3 \int_0^r [1 + \xi(r^\prime, a)]r^{\prime 2}\mathrm{d} r^\prime,
\end{equation}
where $\bar{n}(t)$ is the mean object number density at time $t$ and $\xi(r,a)$ is the two-point correlation function at scale factor $a$ \cite{2010gfe..book.....M}. The conservation of mass can be expressed as
\begin{equation}
\label{eq:mass_conservation}
    \frac{\partial N(r, t)}{\partial t}  + 4\pi \bar{n}a^2 r^2 [1 + \xi(r,a)] v_{12} = 0. 
\end{equation}
Combining Eq.~(\ref{eq:neighbors}) and Eq.~(\ref{eq:mass_conservation}), we have 
\begin{equation}
\label{eq:full_v12}
    v_{12}(r, a) = - \frac{H(a) a^2 }{[1 + \xi(r, a)]r^2} \frac{\partial}{\partial a} \int_0^r \xi(r^\prime, a)r^{\prime 2}\mathrm{d} r^\prime.
\end{equation}
For large separation $r$ where $\xi(r, a) \ll 1$, $\xi(r, a)$ can be expanded to the first order as $\xi(r, a) = \xi(r, 1) D^2(a)$, with $D(a)$ the linear growth factor. Hence, Eq.~(\ref{eq:full_v12}) is written as 
\begin{equation}
    v_{12}(r, a) = -\frac23 Hra f(a)  \frac{3 \int_0^r \xi(r^\prime,a)r^{\prime 2}\mathrm{d} r^\prime}{r^3 \left[1 + \xi(r, a)\right]},
\end{equation}
where $f(a)\equiv \mathrm{d}\ln{D} /\mathrm{d}\ln{a}$ is the linear growth rate. To good approximation, $f(a ) \approx \Omega_m^{0.55}$ \cite{Linder2005}.

\begin{figure}[tpb]
    \centering
    \includegraphics[width=0.9\linewidth]{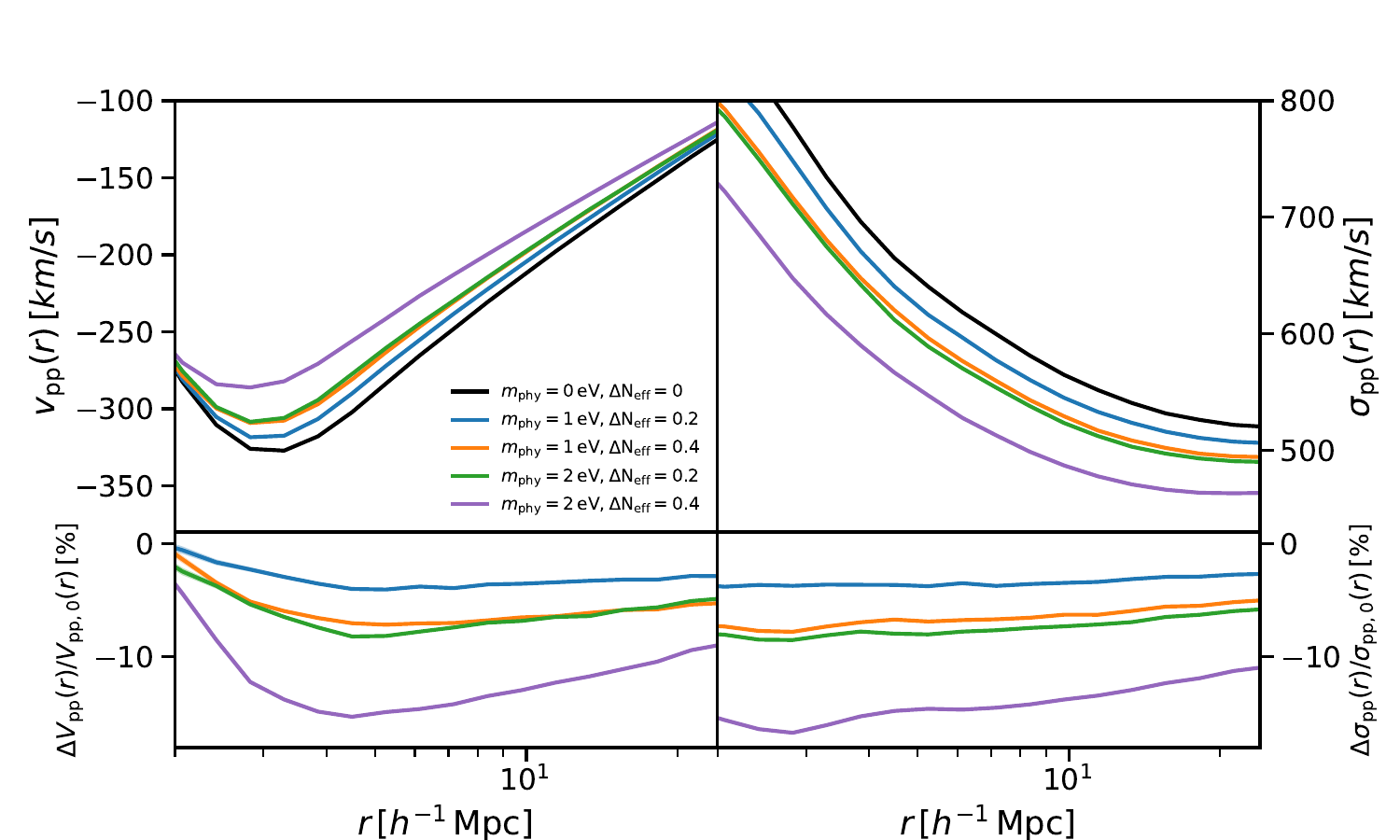}
    \caption{Particle-particle pairwise velocities (upper left) and dispersions (upper right) for different sterile neutrino parameter combinations and their fractional deviations with respect to A0 (lower panels).}
    \label{fig:ppwv_ratio}
\end{figure}

Figure~\ref{fig:ppwv_ratio} displays the particle-particle pairwise velocity $v_{\rm pp}$ and dispersion $\sigma_{\rm pp}$ as well as their corresponding fractional deviations from the fiducial case, $(v_{\rm pp} - v_{\rm pp, 0})/v_{\rm pp, 0}$ and $(\sigma_{\rm pp}-\sigma_{\rm pp, 0})/\sigma_{\rm pp, 0}$, with uncertainties staying within $1\%$. Both $v_{\rm pp}$ and $\sigma_{\rm pp}$ decrease in magnitude for larger $m_{\rm phy}$ or $\Delta N_{\rm eff}$, showing that particles are less attracted by each other. The deviations of pairwise velocity from that of A0 can be well parameterized by $m_{\rm eff}\equiv m_{\rm phy}\cdot \Delta N_{\rm eff}$. For example, B2 and C1, which have the same value of $m_{\rm eff}$, show similar $v_{\rm pp}$ and $\sigma_{\rm pp}$.

\begin{figure}[tpb]
    \centering
    \includegraphics[width=0.9\linewidth]{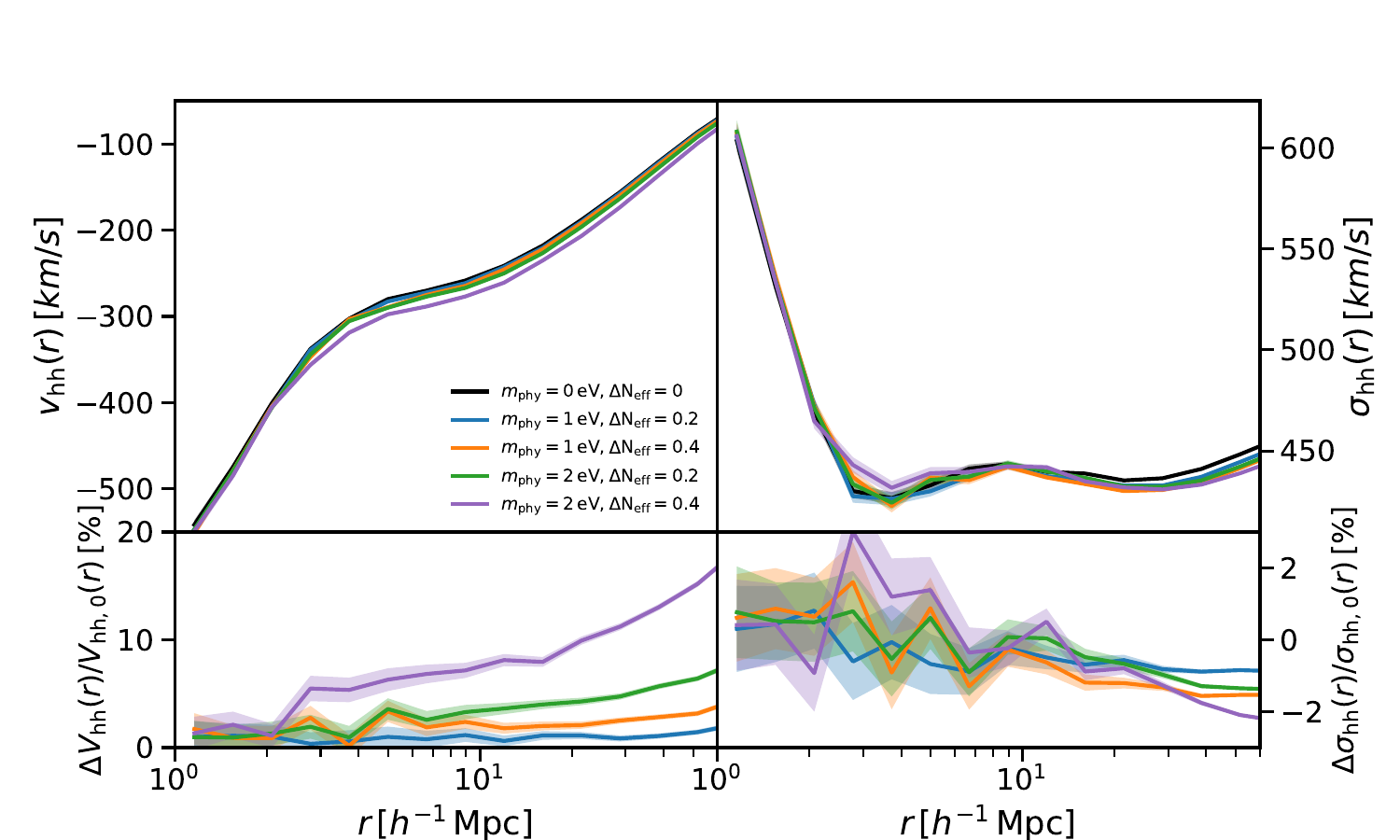}
    \caption{Same as Figure~\ref{fig:ppwv_ratio}, but for halo pairs.}
    \label{fig:hppwv}
\end{figure}

We select halos in the mass range of $[10^{13}, 10^{14}]\,\mathrm{Mpc}\,h^{-1}$ to calculate the halo-halo pairwise velocity $v_{\rm hh}$ and dispersion $\sigma_{\rm hh}$. Here, the distinct host halos are considered, which contain at least one bound halo as the main halo located at the center, with each halo composing of at least 200 CDM particles. The halo mass $M_{\rm 200c}$ is defined as the mass of CDM particles contained within a radius $R_{\rm 200c}$ enclosing a mean over-density of 200 times the critical density. Figure~\ref{fig:hppwv} displays $v_{\rm hh}$ and $\sigma_{\rm hh}$ their fractional deviations from those of the fiducial case A0, $R^v_{\rm hh} = (v_{\rm hh} - v_{\rm hh,0})/v_{\rm hh,0}$ and $R^\sigma_{\rm hh} = (\sigma_{\rm hh} - \sigma_{\rm hh,0})/\sigma_{\rm hh,0}$. For $r>4\, \mathrm{Mpc}\,h^{-1}$, $|v_{\rm hh}|$ increases as $m_{\rm phy}$ or $\Delta N_{\rm eff}$ increases, while $\sigma_{\rm hh}$ decreases. 
Similar to the total matter power spectrum, we take the averaged fractional deviations $\bar{R}_{\rm hh}^{v/\sigma}$ of $v_{\rm hh}$ and $\sigma_{\rm hh}$ over the range $[6, 20]\, \mathrm{Mpc}\, h^{-1}$, and provide a fitting formula:
\begin{equation}
    \bar{R}_{\rm hh}^{v/\sigma}(m_{\rm phy}, \Delta N_{\rm eff}) = C_{nn}^{v/\sigma} \cdot (\Delta N_{\rm eff})^2 +  C_{mn}^{v/\sigma} \cdot \frac{m_{\rm phy}}{1\,\rm{eV}} \times \Delta N_{\rm eff} + C_n^{v/\sigma} \cdot \Delta N_{\rm eff}.
\end{equation}
The fitted coefficients are presented in Table~\ref{tab:fitting formula}.
\begin{table}[tpb]
    \centering
    \caption{Fitting results of $\bar{R}_{\rm hh}^{v}$ and $\bar{R}_{\rm hh}^{\sigma}$} for $M_h \in [10^{13}, 10^{14}]\,\mathrm{M}_\odot\,h^{-1}$ at the range $[6,20]\,\mathrm{Mpc}\,h^{-1}$.
    \label{tab:fitting formula}
    \begin{tabular}{cccc}
        \toprule
        quantity & $C_{nn}^{v/\sigma}$ & $C_{mn}^{v/\sigma}$ & $C_{n}^{v/\sigma}$ \\
        \midrule
        $v_{\rm hh}$ & $0.034\pm0.027$ & $0.134\pm0.006$ & $-0.097\pm0.012$ \\ 
        $\sigma_{\rm hh}$ & $0.041\pm0.005$ & $0.014\pm0.001$ & $-0.052\pm0.002$ \\
    \bottomrule
    \end{tabular}
\end{table}

\subsection{Effects on halo mass function and halo velocity function}
\label{subsec:hmf}
We have shown that CDM structures are impacted by the presence of sterile neutrinos through the suppression and enhancement of the matter power spectrum and pairwise velocity, respectively. On the other hand, the statistics of CDM halos can also be used to constrain models. In this section, we investigate the impact of sterile neutrinos on the CDM halo statistics. 

\begin{figure}[tpb]
    \centering
    \includegraphics[width=0.8\linewidth]{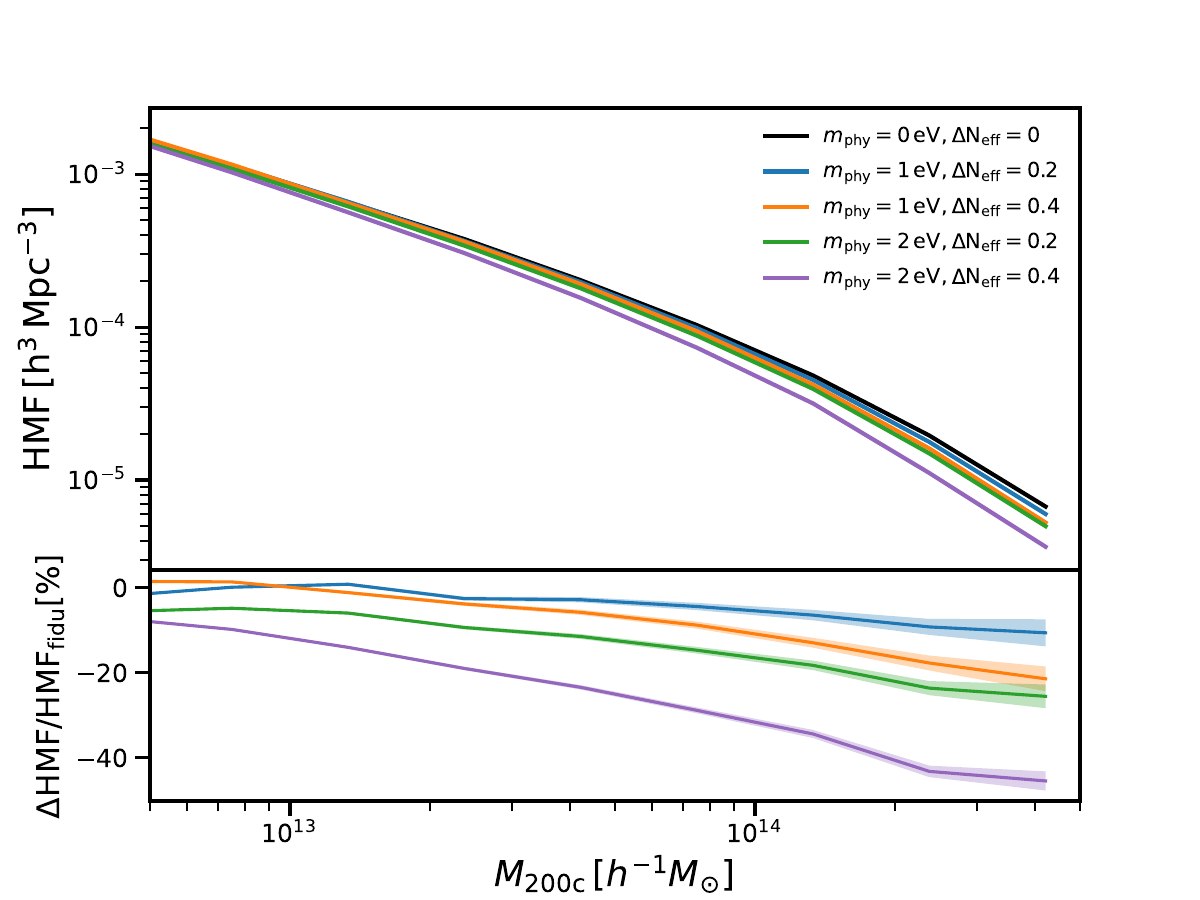}
    \caption{Same as Figure~\ref{fig:pk_effect}, but for halo mass function.}
    \label{fig:HMF}
\end{figure}

Now we will discuss the HMF for different sterile neutrino parameters. The HMF shows the mass distribution of dark matter halos, giving the number density of halos $n$ in each mass bin. We parameterize it in the following way: 
\begin{equation}
     \mathrm{HMF}(M_{200c})\equiv \frac{\mathrm{d}n}{\mathrm{d}\log M_{200c}}=T(\sigma, z)\frac{\rho}{M_{200c}} \frac{\mathrm{d}\log \sigma^{-1}(M, z)}{\mathrm{d}M_{200c}},
\end{equation}
where $\frac{\mathrm{d}n}{\mathrm{d}\log M_{200c}}$ is the comoving number density of dark matter halos per unit mass (in log basis) at redshift $z$, $\rho$ is the mean density of the CDM and baryons, and
\begin{equation}
    \sigma^2 = \frac{1}{2\pi^2} \int_0^\infty k^2 P(k,z) W^2 (k, R_{200c}) \mathrm{d} k,
\end{equation}
with $P(k,z)$ being the total matter power spectrum at $z$  and $W(k, R_{200c})$ the top-hat window function with radius $R_{200c}$. $T(\sigma, z)$ is a factor that can be well-fitted by the Sheth-Tormen approximation in the $\Lambda$CDM model \cite{Sheth_1999}. Figure~\ref{fig:HMF}  displays the HMFs for different sterile neutrino parameter combinations. Despite the uncertainties, the HMF is reduced for increasing $m_{\rm phy}$ or $\Delta N_{\rm eff}$, and the reduction increases at the high-mass end as expected, which is similar to the suppression found with active neutrinos \cite{MTNG}. For instance, the suppression of the HMF is up to 40-50\% for $m_{\rm phy} = 2\, \mathrm{eV}$ and $\Delta N_{\rm eff} = 0.4$ compared to the fiducial case A0, showing the sterile neutrino effect to delay the structure formation or smooth out structures. Moreover, different combinations of $m_{\rm phy}$ and $\Delta N_{\rm eff}$ with the same $m_{\rm eff}$ (B2 and C1) exhibit different strengths of reduction.

\begin{figure}[tpb]
    \centering
    \includegraphics[width=0.85\linewidth]{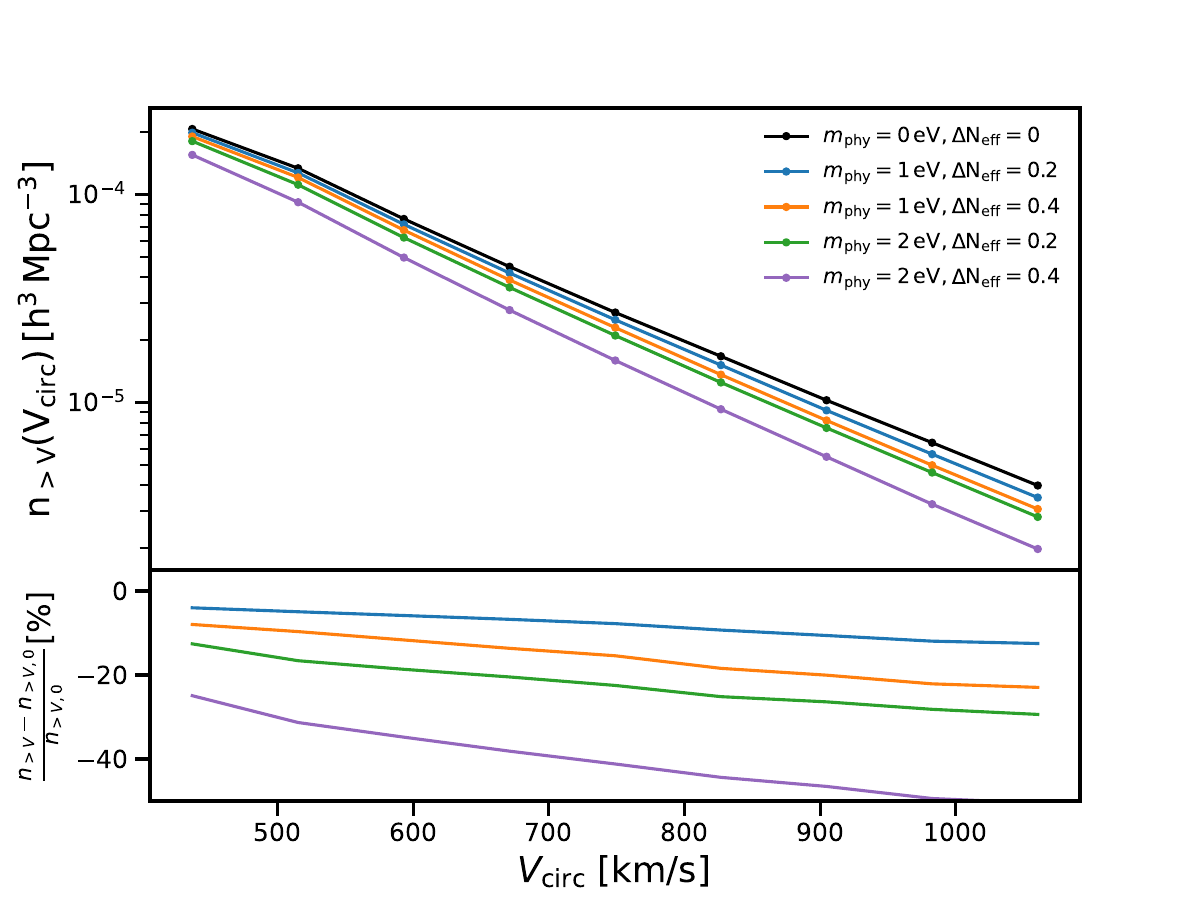}
    \caption{Same as Figure~\ref{fig:pk_effect}, but for cumulative halo velocity function.}
    \label{fig:HVF}
\end{figure}

The maximum circular velocity is another tool to characterize halos \cite{Conroy2006, Klypin2011, Trujillo_Gomez_2011}, which is defined as 
\begin{equation}
    V_{\rm circ} =\left. \sqrt{\frac{GM_{<r}}{r}}\right\vert_{\mathrm{max}} ,
\end{equation}
where $M_{<r}$ is the enclosed mass within radius $r$ and ``max" indicates the maximum circular velocity. The MCV is reached in the central halo region, which is expected to correlate better with the stellar or luminous component and be less sensitive to tidal stripping.  Figure~\ref{fig:HVF} displays the cumulative velocity function, $n_{>V}(V_{\rm circ})$, which represents the number density of dark matter halos with a circular velocity larger than $V_{\rm circ}$. The cumulative velocity function is reduced for larger $m_{\rm phy}$ and $\Delta N_{\rm eff}$. This reduction increases with $V_{\rm circ}$ and reaches around 50\% at the high-speed end for the case C2. Compared to the HMF, the parameter degeneracy (between B2 and C1) is broken more for the cumulative velocity function. 

\section{Conclusions}
\label{sec:con}
In this paper, we present the sterile neutrino effects on the matter power spectrum, pairwise velocity, and dark matter halo statistics calculated using sterile neutrino-involved cosmological simulations. The correction of gravitational potential due to sterile and active neutrinos is calculated through the PM grid method. The sterile neutrino over-density responds to the nonlinear CDM plus baryon density according to the linear theory. This evolution equation may also describe other decoupled particles whose over-densities are small. For instance, similar to the neutrinos, the cosmic QCD axions with eV masses have significant thermal velocities and their over-densities are also small. More interestingly, they may have a Fermi-Dirac-shaped distribution through thermal production after decoupling \cite{Notari:2022ffe}. In this case, the eV-scale QCD axions can also be studied using this cosmological simulation. The simulations converge well with respect to resolutions.

To consistently investigate the impact of sterile neutrinos on the large-scale structure, we run the simulations with cosmological parameters refitted with the Planck CMB and BAO data for each set of $\{m_{\rm phy}, \Delta N_{\rm eff}\}$. Our key conclusions are summarized below.
\begin{enumerate}[(i)]
    \item For the large-scale spatial information, the total matter power spectrum and two-point correlation function are studied. Both of these are suppressed by up to around $40\%$ for $m_{\rm eff}=0.8\,\rm eV$ compared to the fiducial case with no sterile neutrinos. The degeneracy between effects of $m_{\rm phy}$ and $\Delta N_{\rm eff}$ is broken for $k \gtrsim 1 h\,\mathrm{Mpc}^{-1}$. A fitting formula for the averaged fractional deviations of total matter power spectrum is presented.
    \item We calculate the halo-halo pairwise velocity and its dispersion, showing the tendency for objects to be attracted by each other. The halo-halo pairwise velocity (dispersion) increases (decreases) in magnitude by up to $15\%$ ($2\%$) as $m_{\rm phy}$ or $\Delta N_{\rm eff}$ increases for halos in the mass range $[10^{13}, 10^{14}]\,\mathrm{M_\odot}\,h^{-1}$. Similar to the matter power spectrum, we present the fitting formulae for the averaged fractional deviations of $v_{\rm hh}$ and $\sigma_{\rm hh}$.
    \item The presence of sterile neutrinos suppresses the halo mass and cumulative circular velocity functions by up to $40$-$50\%$ for $m_{\rm eff}=0.8\,\rm eV$. These fractional deviations are larger at their high-mass and high-speed ends.
\end{enumerate}

\appendix
\section{Grid-based sterile neutrino-involved N-body simulations}
\subsection{Linear evolution of sterile neutrino over-density}
\label{Appen: evolution}
This section briefly overviews the linear evolution of sterile neutrino over-density, following previous neutrino studies \citep{Ali12, Zeng:2018pcv, Wong:2021ats, Zhang:2023otn}. 
Starting with the Vlasov equation for the sterile neutrinos, 
\begin{equation}
\label{eq:vlasov}
    \frac{\mathrm{d}F_{\nu_s}}{\mathrm{d}t} =  \frac{\partial F_{\nu_s}}{\partial t} + \mathbf{\dot{r}} \cdot \frac{\partial F_{\nu_s}}{\partial \mathbf{r}} + \mathbf{\dot{v}} \cdot \frac{\partial F_{\nu_s}}{\partial \mathbf{v}} = 0, 
\end{equation}
where $F_{\nu_s}(\mathbf{r}, \mathbf{v})$ is the sterile neutrino distribution at position $\mathbf{r}$. In the non-relativistic regime, the acceleration $\mathbf{\dot{v}}$  can be obtained through
\begin{equation}
\label{eq:poisson}
    \mathbf{\dot{v}} = \mathbf{\nabla} \Phi  = - G \int \rho_t \frac{\mathbf{r} - \mathbf{r^\prime}}{|\mathbf{r} - \mathbf{r^\prime}|^3} \mathrm{d^3 r^\prime},
\end{equation}
where $\rho_t (r, t)$ is the total energy density of all matters, including cosmological neutrinos and sterile neutrinos, baryons, and cold dark matter. Here, one can divide $F_{\nu_s}$ into an unperturbed initial distribution $f_{\nu_s}^0$  plus a gravitational distorted term which is position dependent, $f_{\nu_s}^1 (\mathbf{r}, \mathbf{v})$, so that, 
\begin{equation}
\label{eq:linearization}
    F_{\nu_s} = f_{\nu_s}^0(\mathbf{v}) + f_{\nu_s}^1(\mathbf{r},\mathbf{v}).
\end{equation}
For simplicity, we transform the physical coordinates $(t, \mathbf{r}, \mathbf{v})$ into the super-comoving coordinates $(s, \mathbf{x}, \mathbf{v})$ defined as: 
\begin{equation}
\label{eq: coordinates}
    \mathrm{d} s = \frac{\mathrm{d} t}{a^2(t)}, \quad \mathbf{x} = \frac{\mathbf{r}}{a(t)}, \quad \mathbf{u}\equiv \frac{\mathrm{d} \mathbf{x}}{\mathrm{d}s} = a(t) \mathbf{v} - Ha(t) \mathbf{r}.
\end{equation}
The Vlasov equation can be rewritten as 
\begin{equation}
\label{eq:v1}
    \frac{1}{a^2} \frac{\partial F_{\nu_s}}{\partial s} + \frac{\mathbf{u}}{a^2}\cdot \frac{\partial F_{\nu_s}}{\partial \mathbf{u}}  - \ddot{a}a\mathbf{x} \cdot \frac{\partial F_{\nu_s}}{\partial \mathbf{u}} + a \mathbf{v} \cdot \frac{\partial F_{\nu_s}}{\partial \mathbf{u}} = 0.
\end{equation}
Combining Eq.~(\ref{eq:poisson}),  the Friedmann equation, and Eq.~(\ref{eq:linearization}), Eq.~(\ref{eq:v1}) can be linearized as
\begin{equation}
    \frac{\partial f_{\nu_s}^1}{\partial s} + \mathbf{u} \cdot \frac{\partial f_{\nu_s}^1}{\partial \mathbf{x}} - G a^4  \frac{\partial f_{\nu_s}^0}{\partial \mathbf{u}}\cdot \int \bar{\rho} _t\delta_t(s, \mathbf{x^\prime}) \frac{\mathbf{x}-\mathbf{x^\prime}}{|\mathbf{x}-\mathbf{x^\prime}|^3}\mathrm{d^3 x^\prime},
\end{equation}
where $\bar{\rho}_t\delta_t \equiv \rho_t - \bar{\rho}_t = \bar{\rho}_{cb}  \delta_{cb} + \bar{\rho}_\nu\delta_\nu$. By applying the Fourier transformation and integrating out $s$ from initial time $s_i$, one can obtain 
\begin{equation}
\begin{aligned}
    \tilde{f}_{\nu_s}^1(s, \mathbf{k}, \mathbf{u}) + \int_{s_i}^s e^{-i \mathbf{k\cdot u}(s-s^\prime)}  4\pi G a^4 \frac{i\mathrm{k}}{\mathrm{k^2}} \frac{f^0 _{\nu_s}}{\partial \mathbf{u}}\left[ \bar{\rho}_{cb}(s^\prime) \tilde{\delta}_{cb} (s^\prime, \mathbf{k}) + \bar{\rho}_{\nu_s}(s^\prime) \tilde{\delta}_{\nu_s} (s^\prime, \mathbf{k}) \right]\mathrm{d}s^\prime = \\
    \tilde{f}_{\nu_s}^1(s, \mathbf{k}, \mathbf{u}) e^{-i \mathbf{k\cdot u}(s-s_i)},
\end{aligned}
\end{equation}
where '$\sim$'denotes the corresponding Fourier transformed variables.  Following Ali-Haimoud's work \cite{Ali12}, the initial $\tilde{f}_{\nu_s}^1 (s_i, \mathbf{k}, \mathbf{u})$ can be expanded by Legendre polynomials 
\begin{equation}
    \tilde{f}_{\nu_s}^1(s_i, \mathbf{k}, \mathbf{u}) = \sum_{l=0 }^{\infty } i^l  \tilde{f}_{\nu_s}^{1, (l)}(s_i, \mathbf{k}, \mathbf{u}) P_l(\hat{k}\cdot\hat{u}),
\end{equation}
with the coefficients approximated as 
\begin{equation}
    \begin{aligned}
        &\tilde{f}_{\nu_s}^{1, (0)} = f_{\nu_s }^0 \tilde{\delta}_{\nu_s} (s_i, \mathbf{k}),\\
        &\tilde{f}_{\nu_s}^{1, (1)} = \frac{\mathrm{d}f_{\nu_s}^0(u)}{\mathrm{d}u} k^{-1} a_i \tilde{\theta}_{\nu_s} (s_i, \mathbf{k})=-\frac{\mathrm{d}f_{\nu_s}^0(u)}{\mathrm{d}u} k^{-1} a_i^2 H(a_i)\tilde{\delta}_{\nu_s}(s_i, \mathbf{k}),\\
       & \tilde{f}_{\nu_s}^{1, (l)} = 0 (l\geq2),
    \end{aligned}
\end{equation}
where $\tilde{\theta}_{\nu_s}(s_i, \mathbf{k})$ is the sterile neutrino velocity divergence in $k$ space at initial time. After integrating out $\mathbf{u}$, the evolution of sterile neutrino over-density $ \delta_{\nu_s}$ follows
\begin{equation}
\label{eq:evolution}
\begin{aligned}
    \tilde{\delta}_{\nu_s} = \tilde{\delta}_{\nu_s}(s_i, \mathbf{k})\Phi\left[\mathbf{k}(s-s_i)\right] \left[1 + (s - s_i)a_i^2 H(a_i)\right] + 4\pi G \int_{s_i}^s a^4 (s-s^\prime) \Phi\left[ \mathbf{k}(s-s^\prime)\right] \\ 
    \times \left[ \bar{\rho}_{cb}(s^\prime) \delta_{cb} (s^\prime, \mathbf{k}) + \sum_{i} \bar{\rho}_{\nu_i}(s^\prime) \delta_{\nu_i} (s^\prime, \mathbf{k}) + \bar{\rho}_{\nu_s}(s^\prime) \delta_{\nu_s} (s^\prime, \mathbf{k}) \right] \mathrm{d}s^\prime,
\end{aligned}
\end{equation}
where $\Phi(\mathbf{q}) \equiv \frac{\int f^0_{\nu_s} e^{-i\mathbf{q\cdot u}}\mathrm{d^3 u}}{\int f^0_{\nu_s}\mathrm{d^3 u}}$. 

\subsection{Grid-based Method}
\label{append:simulation}
Here, the detailed procedure of the simulations is overviewed:
\begin{enumerate}
    \item The Boltzmann equation solver \texttt{CAMB} is used to generate the initial power spectrum of CDM plus baryons and neutrinos. We modify the calculation of the neutrino energy density and pressure so that the mass of neutrinos can be input directly.
    \item We generate the initial power spectrum of CDM plus baryons $P_{\rm cb}(k)$ and $3+1$ neutrinos $P_\nu(k)$ at redshift z = 99. $R_{\rm ini}(k)\equiv P_{\rm cb}(k)/P_\nu(k)$ is also recorded. In \texttt{2LPTic}, $P_{\rm cb}(k)$ is used to generate the initial conditions for our cosmological simulation. We have modified the cosmology so that the effect of neutrinos is included. 
    \item At the $0$th time step of the cosmological simulation, $P_{\rm cb}(k)$ are calculated with proper binning $\{k_i\}$, so that the initial power spectrum of $3+1$ neutrinos are calculated by $P_\nu(k)= P_{\rm cb}(k)R_{\rm ini}(k)$. Then, $\tilde{\delta}_{\nu_i}$ and $\tilde{\delta}_{\nu_s}$ estimated from $\tilde{\delta}_\nu (k)= \left(P_\nu(k)\right)^{1/2}$ are saved as the initial conditions of Eq.~(\ref{eq:evolution}).
    \item At the $n$th time step, the tables of $\tilde{\delta}_{\nu_i} (s, k)$ and $\tilde{\delta}_{\nu_s} (s, k)$ are made for interpolation so that the integration of Eq.~(\ref{eq:evolution}) can be calculated. Then the correction of gravitational potential is obtained by $\Phi =  4\pi G \tilde{\rho}_{\rm t}(\mathbf{k})$ using Eq.~(\ref{eq:correction of density}). Hence, the long-range force on CDM simulation particles includes the neutrino free-streaming effect. 
    \end{enumerate}

\section{Refitting of cosmological parameters: different sterile neutrino parameters}
\begin{figure}[tpb]
    \centering
    \includegraphics[width=0.9\linewidth]{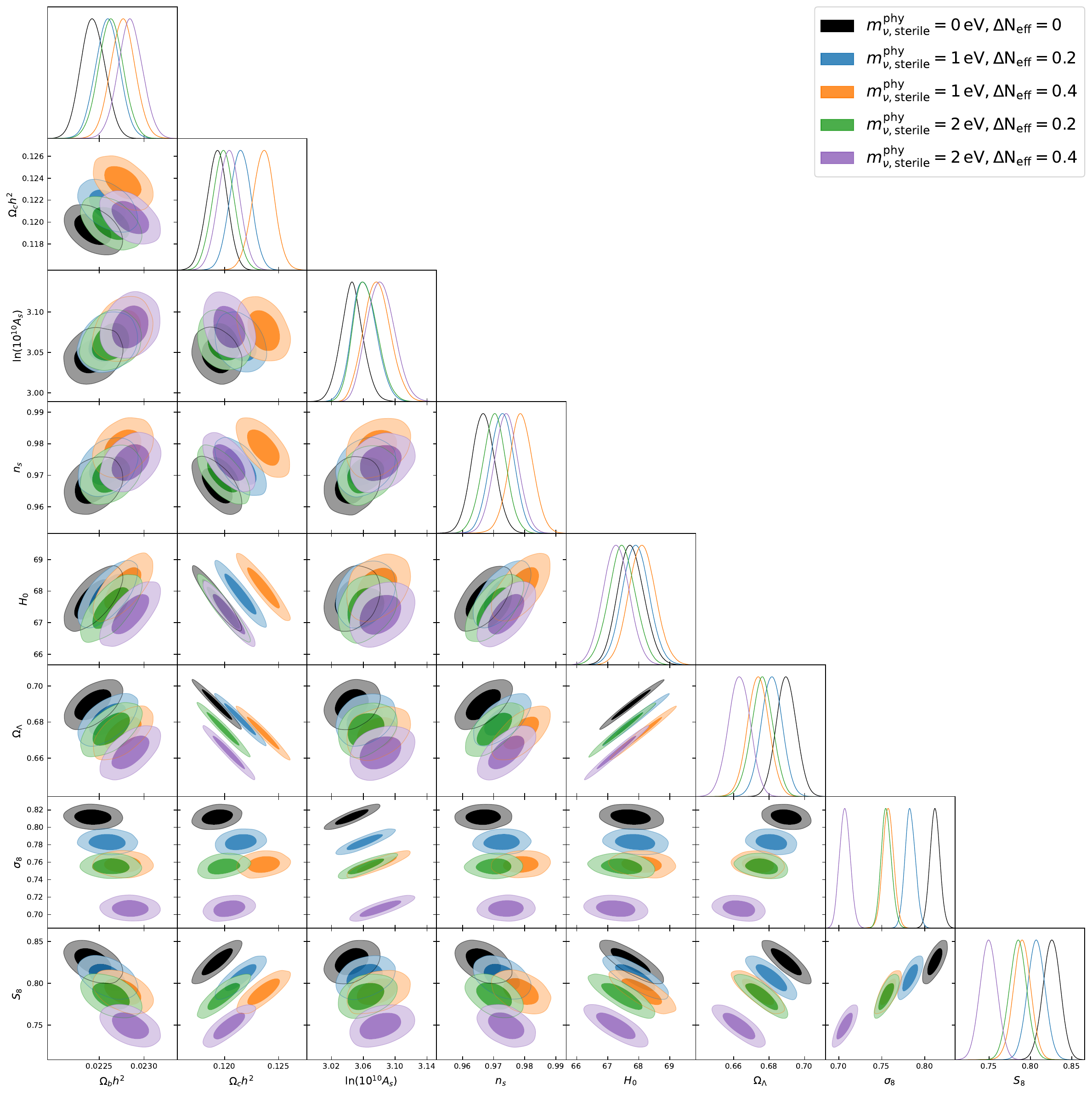}
    \caption{1D posterior PDFs and 2D contours of selected cosmological parameters obtained by MCMC refitting of Planck data and BAO data with different sterile neutrino parameters (shown by colors). The lighter and darker shadow regions show the 2D contour for 95\% C.L. and 68\% C.L., respectively. }
    \label{fig:MCMC}
\end{figure}

Figure~\ref{fig:MCMC} displays the MCMC refitting results by Planck 2018 plikHM\_TTTEEE and BAO data for fixed $m_{\rm phy}$ and $\Delta N_{\rm  eff}$ \cite{planck18, BAO}. With a fixed $\Delta N_{\rm eff}$, increasing $m_{\rm phy}$ results in lower $\Omega_\nu$ and decreases the Hubble parameter $H_0$. If one fixes $m_{\rm eff}$, or sterile neutrino energy density, (see the red and gray contours in Figure~\ref{fig:MCMC}), increasing $\Delta N_{\rm eff}$ indeed alleviates the Hubble tension. It is shown that both $\Omega_m$ and $\sigma_8$ are only sensitive to $m_{\rm eff}$. One can alleviate the $S8$ tension by increasing $m_{\rm eff}$.

\section{Cosmic variance}

This paper uses a random seed value of $667788$ for our simulations. Additionally, we run simulations with a different random seed value of $271828$ to evaluate the stability of the sterile neutrino effect. Here, we take the fractional deviation of the matter power spectrum as an example.

\begin{figure}[tpb]
    \centering
    \includegraphics[width=0.85\linewidth]{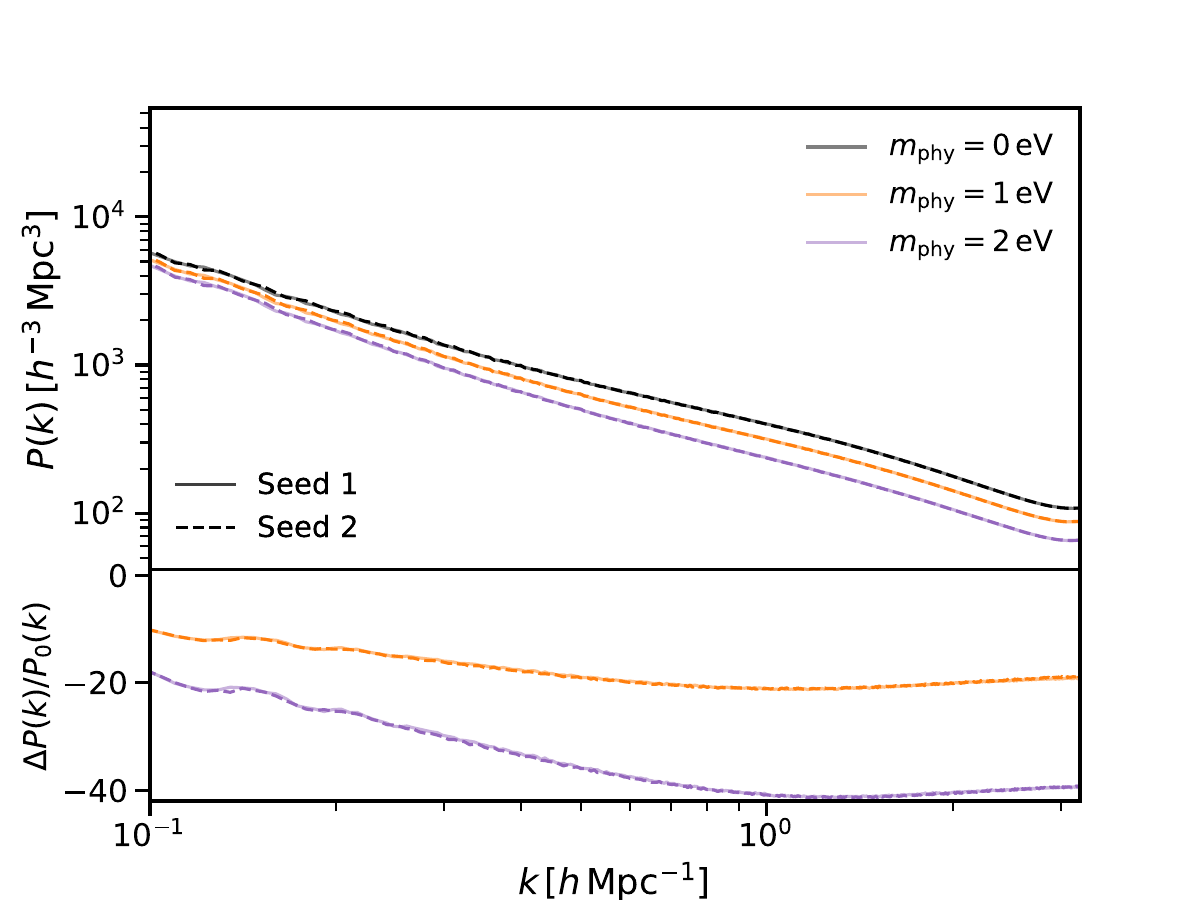}
    \caption{Total matter power spectra (upper panel) and their fractional deviations (lower panel) with respect to the fiducial A0 case for different $m_{\rm phy}$ with two different random seeds: 667788 (solid lines) and 271828 (dashed lines). A fixed value of $\Delta N_{\rm phy}=0.4$ is used for all curves.}
    \label{fig:cosmic_variance}
\end{figure}

Figure~\ref{fig:cosmic_variance} shows the results with solid and dashed lines representing the random seeds 667788 and 271828, respectively. we find the fractional deviations with respect to their corresponding fiducial A0 case show only small cosmic variance. Such stability can be also found when varying $\Delta N_{\rm eff}$ with fixed $m_{\rm phy}$. Consequently, this paper uses the random seed value of 667788. 

\acknowledgments

The computational resources used in this work were kindly provided by the Chinese University of Hong Kong Central Research Computing Cluster. Furthermore, this research is supported by grants from the Research Grants Council of the Hong Kong Special Administrative Region, China, under Project No.s AoE/P-404/18 and 14300223. All plots in this paper are generated by \texttt{Matplotlib} \cite{Hunter:2007}. We also use \texttt{NumPy} \cite{Harris:2020xlr}, \texttt{SciPy} \cite{Virtanen:2019joe} and \texttt{Pandas} \cite{McKinney:2010nts}.

\end{document}